# Non-Stationary Star and the Trajectory of a Circulating Test Body

Walter Petry
Mathematisches Institut der Universitaet Duesseldorf, D-40225 Duesseldorf
E-mail: wpetry@meduse.de
petryw@uni-duesseldorf.de

Abstract: A simple model of a spherically symmetric, pulsating star is calculated. The application to th Sun gives a 166-min radial pulsation. The theory of gravitation in flat space-time implies for a spherically symmetric, non-stationary star small time-dependent exterior gravitational effects. The perturbed equations of motion of a test body moving around the non-stationary star are given. The test body moves away from the center during the epoch of collapsing star and moves towards the center during the epoch of expanding star but the converse is also possible under some conditions. The application to the Sun-Earth system is too small to be measured. This effect may be measurable for very compact, non-stationary objects circulated of a nearby test body.

## 1. Introduction

A simple model of a radially pulsating star is calculated. The application to the Sun gives a 166-min radial pulsation in good agreement with the measured period of 160 min (see [1-4]). A different explanation of this effect has been given in the e-print [5] with a maximal period of 148 min.

The theory of gravitation in flat space-time gives to 2-post Newtonian approximation of a non-stationary, spherically symmetric star small time-dependent gravitational effects in the exterior of the body. This is different to the general relativity theory by virtue of the theorem of Birkhoff which implies a stationary field in the exterior of the star. The perturbed equations of motion of a test body in this non-stationary gravitational field are studied. The orbit oscillates around a circle with the same frequency as the pulsating star but with a very small amplitude. The test body moves towards the central body during the epoch of expanding star and moves away from the central body during the epoch of its contraction but the converse may also arise under suitable conditions. The deviation of the orbit of a test body in the gravitational field of the pulsating star of the Sun-Earth system is too small to be measured. The oscillation of the trajectory of a test body may be measurable for very compact, non-stationary objects with a nearby ciculating test body.

The starting point is the theory of gravitation in flat space-time [6]. This gravitational theory gives for all the well-known effects the same results as the general theory of relativity to the demanded accuracy, namely: redshift, deflection of light, perihelion, radar time-delay, 1-post-Newtonian approximation, gravitational radiation, and the precession of the spin axis of a gyroscope in the orbit of a rotating body. But there exist also differences to the results of general relativity: (1) The theorem of Birkhoff is not valid. The gravitational field of a non-stationary, spherically symmetric star is time-dependent in the exterior of the body ( see [7,8]). In the present paper this result is applied to the motion of a test body circulating a spherically symmertric star [9]. (2) The gravitational theory in flat space-time gives non-singular, homogeneous, isotropic, cosmological models (see e.g. [10,11 ]).

## 2. Pulsating Star

In this section a non-stationary, spherically symmetric, homogeneous star to Newtonian accuracy is studied. Let us follow along the lines of paper [9]. It holds

$$\frac{\partial v}{\partial t} = -v\frac{\partial v}{\partial r} - \frac{1}{\rho}\frac{\partial pc^2}{\partial r} - \frac{4\pi k}{r^2}\int_0^r x^2 \rho\, dx$$

$$\frac{\partial \Pi}{\partial t} = -v\frac{\partial \Pi}{\partial r} - \frac{pc^2}{\rho}\frac{1}{r^2}\frac{\partial}{\partial r}(r^2 v)$$

$$\frac{\partial \rho}{\partial t} = -\frac{1}{r^2}\frac{\partial}{\partial r}(r^2 \rho v). \qquad (2.1)$$

Here, $\rho, p, \Pi, v$ denote density, pressure, internal energy and radial velocity of the star. Subsequently, the equation of state of a non-relativistic degenerate Fermi gas, i.e.

$$pc^2 = \frac{2}{3}\Pi \rho \qquad (2.2)$$

is used. Furthermore, it is assumed that the star is homogeneous, i.e.

$$\rho = \rho(t). \qquad (2.3)$$

We use the ansatz

$$v(r,t) = \frac{r}{R_0}\bar{v}(t) \qquad (2.4a)$$

$$\Pi(r,t) = \frac{R^2(t) - r^2}{R_0^2}\bar{\Pi}(t) \qquad (2.4b)$$

where $R(t)$ denotes the radius of the star and $R_0$ is a fixed unknown constant. The gravitational mass to the accuracy of Newton is given by

$$M = 4\pi \int_0^{R(t)} x^2 \rho\, dx = \frac{4}{3}\pi\bar{\rho}R^3. \qquad (2.5)$$

It follows by the use of (2.2) to (2.5) and elementary calculations

$$\bar{v} = \frac{R_0}{R}\frac{dR}{dt} \qquad (2.6)$$

$$\bar{\Pi} = \beta c^2 \left(\frac{R_0}{R}\right)^4 \qquad (2.7)$$

where $\beta$ is a constant of integration. Furthermore, the following differential equation

$$\frac{d^2R}{dt^2} = \frac{4}{3}\beta c^2 R_0^2 \frac{1}{R^3} - \frac{kM}{R^2} \qquad (2.8)$$

is received (see [9]). Knowing the solution $R(t)$ of equation (2.8) the functions $\bar{\rho}, \bar{v}, \bar{\Pi}$ are obtained by the relations (2.5), (2.6) and (2.7). Equation (2.8) can be integrated implying

$$\left(\frac{dR}{dt}\right)^2 = C - \frac{4}{3}\beta c^2\left(\frac{R_0}{R}\right)^2 + 2\frac{kM}{R} \qquad (2.9)$$

where $C$ is an arbitrary constant of integration. There exist two different kinds of solutions:

**(1)** $C \geq 0$: the radius $R(t)$ of the star contracts to a positive minimum and then it expands for all times.
**(2)** $C < 0$: the radius $R(t)$ of the star oscillates between a minimum radius $R_1$ and a maximum radius $R_2$. The minimum and the maximum radius are

$$R_1 = \left(kM - \left((kM)^2 - \frac{4}{3}\beta c^2 R_0^2 |C|\right)^{1/2}\right)/|C|$$

$$R_2 = \left(kM + \left((kM)^2 - \frac{4}{3}\beta c^2 R_0^2 |C|\right)^{1/2}\right)/|C|. \qquad (2.10)$$

Relation (2.10) gives

$$\frac{R_1 + R_2}{2} = \frac{kM}{|C|}. \qquad (2.11)$$

Relation (2.11) fixes |C| by the mass and the minimum and maximum radius of the star.
The approximate solution of (2.9) has the form

$$R(t) \approx \frac{R_1 + R_2}{2} - \frac{R_2 - R_1}{2} \cos\left(\left(\frac{kM}{((R_1 + R_2)/2)^3}\right)^{1/2} t\right). \tag{2.12}$$

Hence, the solution describes to Newtonian accuracy a non-singular, spherically symmetric, homogeneous, pulsating star. All these results can be found in paper [9].
The period of oscillation is by equation (2.12)

$$t_p = 2\pi \left(\frac{R_m^3}{kM}\right)^{1/2} \tag{2.13}$$

where

$$R_m = (R_1 + R_2)/2 \tag{2.14}$$

is the mean radius of the star. The application of formula (2.13) gives for the Sun with

$$k = 6.67 \cdot 10^{-8} \left[\frac{cm}{gs^2}\right], M = 2 \cdot 10^{33}[g], R_m = 6.96 \cdot 10^{10}[cm] \tag{2.15}$$

the period of radial oscillation

$$t_p = 9.98 \cdot 10^3 [s] = 166 [\min]. \tag{2.16}$$

This is in good agreement with the experimentally measured value of 160 min although the model of the pulsating star is very simple. A different explanation of this period has recently be given in the e-print [5] receiving an upper bound of 148 min.

### 3. Exterior gravitational field

The exterior gravitatioal field of a spherically symmetric, non-stationary star up to 2-post Newtonian order of flat space-time theory of gravitation has been studied in paper [8]. The potentials in spherical coordinates are

$$g_{11} = 1/f(r,t)$$
$$g_{22} = r^2 / g(r,t)$$
$$g_{33} = r^2 \sin^2(\vartheta) / g(r,t)$$
$$g_{44} = -1/h(r,t)$$
$$g_{ij} = 0, i \neq j \tag{3.1}$$

The theory gives by neglecting expressions of $O(1/r^2)$

$$f \approx g \approx 1 - 2\frac{kM}{c^2 r}$$

$$h \approx 1 + 2\frac{kM}{c^2 r} + \frac{1}{c^6} \frac{4\pi k}{r} \frac{\partial \tilde{h}(r,t)}{\partial t} \tag{3.2}$$

with

$$\tilde{h}(r,t) = -\frac{8}{15} \int_0^\infty x^3 \rho \left(4\frac{pc^2}{\rho} v + v^3 + x\frac{pc^2}{\rho}\frac{\partial v}{\partial x}\right) dx +$$

$$+ 4\pi k \left( \frac{32}{9} \int_0^\infty x^4 \rho \int_\infty^x \rho v d\xi\, dx + \frac{176}{15} \int_0^\infty x^2 \rho v \int_0^x \xi^2 \rho\, d\xi\, dx + \frac{64}{9} \int_0^\infty x\rho \int_0^x \xi^3 \rho v d\xi\, dx \right). \qquad (3.3)$$

### 4. Motion of a test body

Assume that the non-stationary star is circulated of test body with an orbit which is approximately a circle with radius $r_0$. The perturbed radius in the non-stationary gravitational field has the form

$$r = r_0 + \Delta r. \qquad (4.1)$$

Then, the perturbed equations of motion of the test body give to linear approximation (see[9])

$$\frac{d^2}{dt^2}\Delta r = -\frac{kM}{r_0^3}\Delta r - \frac{1}{c^4}\frac{2\pi k}{r_0^2}\frac{\partial \tilde{h}}{\partial t}. \qquad (4.2)$$

A solution to the linear differential equation (4.2) in the interval

$$[t_0, t]$$

is given by the formula

$$\Delta r \approx \frac{1}{c^4}\frac{2\pi k}{r_0^2}\sqrt{\frac{r_0^3}{kM}} \int_{t_0}^t \frac{\partial \tilde{h}}{\partial s} \sin\left(\sqrt{\frac{kM}{r_0^3}}(s-t)\right) ds. \qquad (4.3)$$

Integration by parts two times implies by neglecting expressions of

$$O(1/r_0^3)$$

and neglecting periodic terms

$$\Delta r \approx -\frac{1}{c^4}\frac{2\pi k}{r_0^2} \int_{t_0}^t \tilde{h}(s) ds. \qquad (4.4)$$

Éinstein's general theory of relativity yields by Birkhoff's theorem that the exterior gravitational field of a spherically symmetric, non-stationary star can be given by the Schwarzschild metric. Hence, according to Einstein's theory the functions satisfy

$$\tilde{h} \equiv 0, \Delta r \equiv 0.$$

Now, the results of the chapters 3 and 4 are applied to a non-stationary, spherically symmetric star studied in chapter 2.
It follows from equation (3.3) by the use of (2.2) to (2.7) and (2.9)

$$\tilde{h} \approx \frac{2}{35\pi} M(24kM - CR)\frac{dR}{dt}. \qquad (4.5)$$

Equation (4.4) gives by the use relation (4.5)

$$\Delta r(t) \approx -\frac{2}{35}\left( 48\left(\frac{kM}{c^2 r_0}\right)^2 - \frac{kM}{c^2 r_0}\frac{C}{c^2}\frac{R(t)+R(t_0)}{r_0} \right)(R(t) - R(t_0)). \qquad (4.6)$$

This relation (resp. (4.4) with (4.5)) implies that for not too high positive C the radius r(t) of the trajectory increases if the radius of the star R(t) decreases and conversely r(t) decreases if R(t) increases. But under the condition of a sufficiently high positive C the radius r(t) decreases (increases) if R(t) decreases (increases).

The special case C=0 yields:

$$\Delta r \approx -\frac{96}{35}\left(\frac{kM}{c^2 r_0}\right)^2 (R(t) - R(t_0)). \tag{4.7}$$

Let us now consider a pulsating star, i.e. C<0. Put without loss of generality

$$t_0 = \frac{\pi}{2}\left(\frac{R_m^3}{kM}\right)^{1/2}$$

then equation (2.12) gives

$$R(t_0) \approx \frac{R_1 + R_2}{2} = R_m. \tag{4.8}$$

It follows from (4.6) by the equations (2.12) with C<0, (2.13), (4.8) and the inequality

$$R_2 - R_1 \ll R_1 + R_2$$

the change of the trajectory of the test body

$$\Delta r(t) \approx \frac{10}{7}\left(\frac{kM}{c^2 r_0}\right)^2 (R_2 - R_1)\cos\left(\left(\frac{kM}{R_m^3}\right)^{1/2} t\right) \tag{4.9}$$

in the orbit of the pulsating star.
Let us now consider the Sun-Earth system with the mean distance

$$r_0 \approx 1.495 \cdot 10^{13}\,[cm], \left(c \approx 3\cdot 10^{10}\left[\frac{cm}{s}\right]\right) \tag{4.10a}$$

then the factor

$$\left(\frac{kM}{c^2 r_0}\right)^2 \approx 9.83 \cdot 10^{-17} \tag{4.10b}$$

of formula (4.9) implies that the variation of the radius is much too small to be measured.
But relation (4.7) where C=0 yields that for very compact, non-stationary, spherically symmetric objects which are circulated of nearby test bodies such that the factor in (4.9) is not too small the disturbed trajectory may be measurable.